\documentclass[12pt,a4paper,final]{iopart}
\usepackage{iopams}  
\usepackage[breaklinks=true,colorlinks=true,linkcolor=blue,urlcolor=blue,citecolor=blue]{hyperref}

\usepackage{graphicx,color,rotating}
\newcommand{\op}[1]{\ensuremath{#1}}

\newcommand{\PO}{\ensuremath{\op{P}}}

\newcommand{\VO}{\ensuremath{\op{V}}}

\newcommand{\Vnnn}{\ensuremath{\op{V}_{\mathrm{3N}}}}

\newcommand{\rOV}{\ensuremath{\vec{\op{r}}}}

\newcommand{\nuc}[2]{\ensuremath{{}^{#2}\mathrm{#1}}}

\newcommand{\fm}{\ensuremath{\,\mathrm{fm}}}

\newcommand{\MeV}{\ensuremath{\,\mathrm{MeV}}}
\newcommand{\GeV}{\ensuremath{\,\mathrm{GeV}}}

\newcommand{\Cnnn}{\ensuremath{C_{\mathrm{3N}}}}
\newcommand{\ennn}{\ensuremath{E_{\mathrm{3\max}}}}

\newcommand{\symboldiamond}[1][black]{{\color{#1}\scriptsize\begin{turn}{45}$\blacksquare$\end{turn}}}
\newcommand{\symboltriangle}[1][black]{{\color{#1}$\blacktriangle$}}
\newcommand{\symbolbox}[1][black]{{\color{#1}\small$\blacksquare$}}
\newcommand{\symbolcircle}[1][black]{{\color{#1}$\bullet$}}

\definecolor{FGViolet}{rgb}{0.61,0.32,0.61}
\definecolor{FGDarkBlue}{rgb}{0,0,0.6}
\definecolor{FGBlue}{rgb}{0,0,0.8}
\definecolor{FGLightBlue}{rgb}{0.2, 0.6, 0.8}
\definecolor{FGGreen}{rgb}{0.2,0.7,0.2}
\definecolor{FGLightGreen}{rgb}{0.4,1,0.4}
\definecolor{FGYellow}{rgb}{1,0.95,0}
\definecolor{FGDarkYellow}{rgb}{1,0.8,0.3}
\definecolor{FGOrange}{rgb}{0.95,0.5,0.1}
\definecolor{FGRed}{rgb}{0.8,0,0}
\definecolor{FGWhite}{rgb}{1,1,1}
\definecolor{FGLightGray}{rgb}{0.8,0.8,0.8}
\definecolor{FGGray}{rgb}{0.5,0.5,0.5}
\definecolor{FGDarkGray}{rgb}{0.3,0.3,0.3}
\definecolor{FGBlack}{rgb}{0,0,0}

\newcommand{\linemediumsolid}[1][black]{\unitlength 0.7ex
  {\color{#1}
  \begin{picture}(6,1)
  \linethickness{0.4mm}
  \put(0,0.5){\line(1,0){6.0}}
  \end{picture}}\nolinebreak
}

\newcommand{\linemediumdashed}[1][black]{\unitlength 0.7ex
  {\color{#1}
  \begin{picture}(6,1)
  \linethickness{0.4mm}
  \put(0,0.5){\line(1,0){1.5}}
  \put(2.2,0.5){\line(1,0){1.5}}
  \put(4.4,0.5){\line(1,0){1.5}}
  \end{picture}}\nolinebreak
}

\newcommand{\linethinsolid}[1][black]{\unitlength0.5ex
  {\color{#1}
  \begin{picture}(6,1)
  \linethickness{0.12mm}
  \put(0,0.5){\line(1,0){6.0}}
  \end{picture}}\ \nolinebreak
}

\newcommand{\linethindashed}[1][black]{\unitlength0.5ex
  {\color{#1}
  \begin{picture}(6,1)
  \linethickness{0.12mm}
  \put(0,0.5){\line(1,0){1.5}}
  \put(2.2,0.5){\line(1,0){1.5}}
  \put(4.4,0.5){\line(1,0){1.5}}
  \end{picture}}\nolinebreak
}

\begin{document}

\title[Giant Resonances based on Unitarily Transformed Interactions]{Giant Resonances based on Unitarily Transformed Two-Nucleon plus Phenomenological Three-Nucleon Interactions}

\author{A. G\"unther}
\address{$^1$Institut f\"ur Kernphysik, Technische Universit\"at Darmstadt,
64289 Darmstadt, Germany}
\author[cor1]{P. Papakonstantinou$^{1,2}$}
\address{$^1$Rare Isotope Science Project, Institute for Basic Science, Daejeon 305-811, Republic of Korea}
\address{$^2$Institut de Physique Nucl\'eaire, IN2P3-CNRS, Universit\'e Paris-Sud, 91406 Orsay, France} 
\ead{\mailto{ppapakon@ibs.re.kr}}
\author[cor2]{R. Roth}
%\email{anneke.guenther@physik.tu-darmstadt.de}
\address{$^1$Institut f\"ur Kernphysik, Technische Universit\"at Darmstadt,
64289 Darmstadt, Germany}
\ead{\mailto{robert.roth@physik.tu-darmstadt.de}}

%\affiliation{Institut f\"ur Kernphysik, Technische Universit\"at Darmstadt,
%64289 Darmstadt, Germany}
%\address{$^2$Rare Isotope Science Project, Institute for Basic Science, Daejeon 305-811, Republic of Korea}
%\address{$^3$Institut de Physique Nucl\'eaire, IN2P3-CNRS, Universit\'e Paris-Sud, 91406 Orsay, France} 
%\affiliation{Institut f\"ur Kernphysik, Technische Universit\"at Darmstadt,
%64289 Darmstadt, Germany}

%\date{\today}

\begin{abstract}
We investigate giant resonances of spherical nuclei on the basis of the Argonne V18 potential after unitary transformation within the Similarity Renormalization Group or the Unitary Correlation Operator Method supplemented by a phenomenological three-body contact interaction. Such Hamiltonians can provide a good description of ground-state energies and radii within Hartree-Fock plus low-order many-body perturbation theory. The standard Random Phase Approximation is applied here to calculate the isoscalar monopole, isovector dipole, and isoscalar quadrupole excitation modes of the $\nuc{Ca}{40}$, $\nuc{Zr}{90}$, and $\nuc{Pb}{208}$ nuclei. 
Thanks to the inclusion of the three-nucleon interaction and despite the minimal optimization effort, a reasonable agreement with experimental centroid energies of all three modes has been achieved. 
% thanks to the inclusion of the three-nucleon interaction.
%A reasonable agreement with experimental centroid energies of all three modes can be achieved thanks to the inclusion of the three-nucleon interaction.
The role and scope of the Hartree-Fock reference state in RPA methods are discussed. 
\end{abstract}


\submitto{\JPG} 
\maketitle

%%%%%%%%%%%%%%%%%%%%%%%%%%%%%%%%%%%%%%%%%%%%%%%%%%%%%%%%%%%%%%%%%%%%%%
%%%%%%%%%%%%%%%%%%%%%%%%%%%%%%%%%%%%%%%%%%%%%%%%%%%%%%%%%%%%%%%%%%%%%%
%%%%%%%%%%%%%%%%%%%%%%%%%%%%%%%%%%%%%%%%%%%%%%%%%%%%%%%%%%%%%%%%%%%%%%
\clearpage

%%%%%%%%%%%%%%%%%%%%%%%%%%%%%%%%%%%%%%%%%%%%%%%%%%%%%%%%%%%%%%%%%%%%%%
%%%%%%%%%%%%%%%%%%%%%%%%%%%%%%%%%%%%%%%%%%%%%%%%%%%%%%%%%%%%%%%%%%%%%%
\section{Introduction}

The most consistent starting point for nuclear structure theory are nuclear Hamiltonians derived from quantum chromodynamics (QCD) in the framework for chiral effective field theory containing two- and three-nucleon interactions \cite{Epelbaum:2005pn,Machleidt:2011zz}. Using these interactions we can employ unitary transformations, e.g. the Similarity Renormalization Group (SRG) or the Unitary Correlation Operator Method (UCOM), to pre-diagonalize the Hamiltonian and to improve the convergence behavior of various many-body approaches. Recently, this approach was applied successfully to light and medium-mass nuclei in the context of the No-Core Shell Model \cite{Jurgenson:2009qs,Roth:2011ar} and in Coupled-Cluster Theory 
and related methods \cite{Roth:2011vt,Binder:2012mk,Hergert:2012nb,Her2013}. 
 
The computational effort, however, limits the applicability of general three-nucleon interactions in the unitary transformation as well as in the application in many-body methods. Furthermore, to provide an appropriate starting point for the investigation of collective excitations in the framework of the Random Phase Approximation (RPA) the interaction has to reproduce experimental ground-state radii reasonably well---this has not yet been achieved on the basis of chiral two- plus three-nucleon interactions beyond the lightest nuclei. As a preparatory step towards the full inclusion of chiral two- and three-nucleon interactions, we follow a more pragmatic approach by using the unitarily transformed Argonne V18 potential \cite{Wiringa:1994wb} supplemented by a phenomenological three-body contact interaction. This allows us to investigate ground-state properties as well as collective excitations throughout the nuclear mass range up to $\nuc{Pb}{208}$.
In a previous paper the influence of phenomenological three-nucleon interactions on the description of ground-state nuclear properties was investigated \cite{Gunther:2010cs} and a good simultaneous description of ground-state energies and radii was achieved. This was not possible with the pure two-body UCOM interaction employed in earlier studies \cite{Roth:2005ah}. We now examine whether dynamical properties, such as collective modes, show a similar quantitative improvement compared to previous work \cite{Paar:2006ua}. 

In this work we apply the standard RPA to study collective excitations of closed-shell nuclei.
We will show that the results obtained with the unitarily transformed Argonne V18 supplemented by a phenomenological three-body contact interaction agree within $20\%$ with the results for traditional phenomenological potentials like the Gogny D1S interaction \cite{Berger:1991zza}.
The characterization of these hybrid Hamiltonians in standard applications is mandatory to provide a well-defined footing for predictive calculations like the study of low-energy dipole transitions \cite{Papakonstantinou:2010ja}, and investigations in the framework of quasi-particle RPA (QRPA) \cite{Hergert:2011eh}.

In Sec.~\ref{sec:formalism}, we present the formalism and discuss briefly some ground-state properties of closed-shell nuclei across the whole nuclear chart on the basis of Hartree-Fock and many-body perturbation theory. 
In Sec.~\ref{sec:rpa}, we study the isoscalar monopole (ISM), isovector dipole (IVD), and isoscalar quadrupole (ISQ) excitation modes of three chosen nuclei, $\nuc{Ca}{40}$, $\nuc{Zr}{90}$, and $\nuc{Pb}{208}$. 
We provide a critical discussion in Sec.~\ref{sec:conclusion}.

%%%%%%%%%%%%%%%%%%%%%%%%%%%%%%%%%%%%%%%%%%%%%%%%%%%%%%%%%%%%%%%%%%%%%%
%%%%%%%%%%%%%%%%%%%%%%%%%%%%%%%%%%%%%%%%%%%%%%%%%%%%%%%%%%%%%%%%%%%%%%
\section{Formalism and ground-state properties 
\label{sec:formalism} } 

The UCOM and the SRG provide two different approaches for generating soft phase-shift equivalent two-body interactions. These two methods have already been discussed extensively (see \cite{Roth:2010bm} and refs. therein). Based on the Argonne V18 potential we will apply four different classes of unitarily transformed interactions in the following: UCOM(SRG), S-UCOM(SRG), SRG, and S-SRG which were introduced in \cite{Gunther:2010cs}. These transformed two-body interactions are supplemented by a simple three-body contact interaction
%%%
\begin{equation}
 \Vnnn=\Cnnn\ \delta^{(3)}(\rOV_1-\rOV_2) \delta^{(3)}(\rOV_1-\rOV_3)
\end{equation}
%%%
with variable strength $\Cnnn$. The m-scheme matrix elements of the contact interaction can be evaluated on-the-fly, which is of great computational advantage. As single-particle basis the eigenstates of the harmonic oscillator are employed. For calculations beyond the mean-field level we have to introduce a regularization of the contact interaction which is achieved by restricting the total oscillator energy of the three-particle state: $(2 n_1+ l_1)+(2 n_2+ l_2)+(2 n_3+ l_3)\leq E_{3\max}$, where $n$ and $l$ are the principal and the angular momentum quantum numbers of the harmonic oscillator states, respectively \cite{Gunther:2010cs}.

%%%%%%%%%%%%%%%%%%%%%%%%%%%%%%%%%%%%%%%%%%%%%%%%%%%%%%%%%%%%%%%%%%%%%%
%%%%%%%%%%%%%%%%%%%%%%%%%%%%%%%%%%%%%%%%%%%%%%%%%%%%%%%%%%%%%%%%%%%%%%
%%\section{Ground-State Properties
%%\label{sec:mbpt}}

For a first characterization of the four different two- plus three-body interactions, the Hartree-Fock (HF) approximation and many-body perturbation theory are used to calculate ground-state energies and charge radii for selected closed-shell nuclei from $\nuc{He}{4}$ to $\nuc{Pb}{208}$. The formal inclusion of the three-body contact interaction in these two methods has been discussed in detail in Ref. \cite{Gunther:2010cs}. The single-particle basis is truncated with respect to the principal oscillator quantum number $e=2n+l$, which is restricted to $e \leq e_{\mathrm{max}}=14$ with an additional constraint for the orbital angular momentum quantum number $l\leq l_{\mathrm{max}}=10$. The oscillator parameter $a_{\mathrm{HO}}$ is chosen for each nucleus separately such that the experimental charge radius is reproduced by a shell-model Slater determinant built from harmonic oscillator single-particle states. We have observed that for these values of $a_{\mathrm{HO}}$ and for $e_{\max}=14$ the HF ground-state energies deviate by less than $0.1\%$ from their minimum with respect to $a_{\mathrm{HO}}$ and that the mean square radius is stable against variations of $a_{\mathrm{HO}}$. Thus the HF energy and radius are well converged for such large values of $e_{\max}$. 

%%%%%%%%%%%%%%%%%%%%%%%%%%%%%%%%%%%%%%%%%%%%%%%%%%%%%%%%%%%
\begin{figure}[b]
\centering
\includegraphics*[width=0.7\columnwidth]{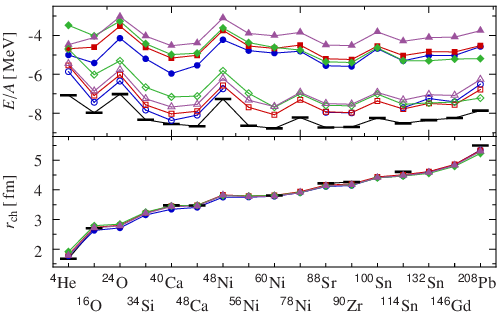}
\vspace{-0.3cm}
\caption{(Color online) Ground-state energies per nucleon and charge radii of selected closed-shell nuclei resulting from HF calculations (filled symbols) and MBPT (open symbols) based on the UCOM(SRG) (\symbolcircle[FGBlue]), S-UCOM(SRG) (\symbolbox[FGRed]), SRG (\symboldiamond[FGGreen]), and S-SRG (\symboltriangle[FGViolet]) interactions using the optimal parameter sets (cf. Tab.~\ref{tab:optparametersets}). Bars indicate experimental values 
\cite{Audi:1995dz,DeJager:1987qc}.}
%\cite{Audi:1995dz,Vries:1987}.}
\label{fig:hf_mbpt}
\end{figure}
%%%%%%%%%%%%%%%%%%%%%%%%%%%%%%%%%%%%%%%%%%%%%%%%%%%%%%%%%%%

%%%%%%%%%%%%%%%%%%%%%%%%%%%%%%%%%%%%%%%%%%%%%%%%%%%%%%%%%%%
\begin{table}
\begin{center}
\begin{tabular}{|l|c|c|}
\hline
& $\alpha\ [\fm^4]$ & $\Cnnn\ [\GeV\fm^6]$ \\
\hline
UCOM(SRG) & 0.16 & 1.6 \\
S-UCOM(SRG) & 0.16 & 2.2 \\
SRG & 0.10 & 4.3 \\
S-SRG & 0.10 & 2.0 \\
\hline
\end{tabular}
\end{center}
\vspace{-0.5cm}
\caption{Optimal parameter sets for the different two- plus three-body interactions.}
\label{tab:optparametersets}
\end{table}
%%%%%%%%%%%%%%%%%%%%%%%%%%%%%%%%%%%%%%%%%%%%%%%%%%%%%%%%%%%

Figure~\ref{fig:hf_mbpt} shows the ground-state energies per nucleon and charge radii for closed-shell nuclei across the whole nuclear chart. The three-body cut-off parameter is set to $\ennn=20$ for all four interactions (cf. \cite{Gunther:2010cs}). The strength of the three-body contact interaction is used to adjust the charge radii to the experimental values.
As observed in Fig.~\ref{fig:hf_mbpt} the strength $\Cnnn$ can be chosen such that the radii are well reproduced across the whole mass range by all four interactions.
This is a remarkable result considering the simplistic structure of the three-body interaction. The optimal values for 
$\Cnnn$ 
%the three-body contact strengths 
are summarized in Table~\ref{tab:optparametersets} together with the corresponding flow parameters.

We point out that, at present, chiral NN+3N interactions still do not provide a good description of the radii beyond the lightest nuclei~\cite{BLC201X}. 
In other words, the systematics shown in Fig.~\ref{fig:hf_mbpt} for such a basic observable cannot be reproduced at present with the most 
advanced chiral Hamiltonians. 
This observation makes the use of a phenomenological three-nucleon correction an appealing option.  

The HF energies reproduce the systematics of the experimental values except for an almost constant shift,
which is due to the missing effects of long-range correlations that cannot be described at the mean-field level. The influence of long-range correlations can be taken into account via second-order many-body perturbation theory (MBPT) \cite{Roth:2005ah}. Here, we apply perturbation theory only to the two-body part of the interaction (cf. \cite{Gunther:2010cs}). The effect on charge radii is negligible, but for the energies the inclusion of the second-order perturbative corrections leads to a substantial improvement (open symbols in Fig.~\ref{fig:hf_mbpt}).
The agreement with the experimental data is not yet perfect, i.e. the differences vary from $0.2$ to $2.4\MeV$ per nucleon, but one has to keep in mind that the energy corrections are not yet fully converged with respect to the single-particle model space size. Furthermore, we only consider the second-order estimate and have no information about the influence of higher orders \cite{Roth:2009up,Langhammer:2012jx}.

In case of the UCOM(SRG) interaction one observes that the energy corrections for some nuclei, mainly the heavier nickel isotopes and $\nuc{Sn}{100}$, do not follow the general trend. The origin of this behavior lies in the corresponding HF single-particle spectra, where some level spacings are collapsed leading to divergent contributions to the perturbative corrections. Therefore, these data points are not shown here.

The dependence of the MBPT results on $E_{3\max}$ when a three-body term is included, especially of the ground state energy, is discussed in Ref.~\cite{Gunther:2010cs}. The calculations of excitation properties discussed next do not depend on $E_{3\max}$.

%%%%%%%%%%%%%%%%%%%%%%%%%%%%%%%%%%%%%%%%%%%%%%%%%%%%%%%%%%%%%%%%%%%%%%
%%%%%%%%%%%%%%%%%%%%%%%%%%%%%%%%%%%%%%%%%%%%%%%%%%%%%%%%%%%%%%%%%%%%%%
\section{Giant Resonances
\label{sec:rpa}} 

For investigating collective excitations we apply the standard Random Phase Approximation (RPA) (cf. e.g. \cite{Paar:2006ua}). In the framework of the HF approximation and the standard RPA the three-body contact interaction is equivalent to the density-dependent two-body interaction \cite{Vautherin:1971aw,Waroquier:1976xz,Bohigas:1978qu}
%%%
\begin{equation}
\VO_{\mathrm{NN}}[\varrho]=\frac{\Cnnn}{6}\ (1+\PO_\sigma)\ \varrho\!\left(\frac{\rOV_1+\rOV_2}{2}\right)\delta^{(3)}(\rOV_1-\rOV_2) \ ,
\end{equation}
%%%
which is used in the RPA implementation for computational reasons. The only difference to the HF plus MBPT calculations discussed 
above 
%%in the previous section 
is the absence of the cut-off $\ennn$ in RPA. The HF calculations providing the basis for RPA are also performed using the density-dependent form of the three-body interaction without the cut-off. As the HF energies are independent of $\ennn$ the chosen implementation has no influence on the RPA results. Although we use density-dependent two-body interaction in RPA we will still refer to it as three-body contact interaction.

Next, we investigate three excitation modes, namely isoscalar monopole (ISM), isovector dipole (IVD), and isoscalar quadrupole (ISQ) excitations, of $\nuc{Ca}{40}$, $\nuc{Zr}{90}$, and $\nuc{Pb}{208}$. The different two- plus three-body interactions introduced 
above 
%%in the previous sections are considered using the parameter sets listed in Table~\ref{tab:optparametersets}. The single-particle basis is again truncated at $e_{\max}=14$. %, $l_{\max}=10$.

As a first benchmark we consider the exhaustion of the classical sum rules. We validate our implementation by using the Gogny D1S interaction, whose momentum dependent terms are of zero range: the ISM and ISQ classical sum rules are then fulfilled within $1\%$ or better. In the case of transformed AV18 interactions, the exhaustion of the classical ISM sum rule lies between $90.5\%$ for $\nuc{Ca}{40}$ calculated with the UCOM(SRG) interaction and $98.4\%$ for $\nuc{Pb}{208}$ calculated with the S-SRG interaction.  The exhaustion of the ISQ sum rule lies between $98.4\%$ for $\nuc{Pb}{208}$ calculated with the SRG interaction and $102.9\%$ for $\nuc{Ca}{40}$ calculated with the UCOM(SRG) interaction.
The deviations from the classical IS sum rules are 
%larger in the monopole than in the quadrupole channel and are 
mostly a consequence of the non-localities of the finite-range interactions used \cite{RS80}. 
%, which are modified by the variation of the flow parameter. 
When including the three-body interaction we use a larger flow parameter in order to compensate for the additional repulsion. The larger flow parameter generates stronger non-localities in the transformed two-body interaction, which in turn influence the exhaustion of the classical sum rules. 
%This effect is revealed most clearly in case of the monopole excitations as this breathing mode involves the whole nucleus whereas the quadrupole excitation can be interpreted as a surface vibration affecting only part of the nucleons.
% The use of a density-dependent interaction and the consequent evaluation of various quantities numerically on a radial grid also introduces some inaccuracies, compared to purely two-body interactions.
It is expected that the Thomas-Reiche-Kuhn sum rule for the IVD mode is significantly enhanced due to the non-localities of the applied interactions. 
Indeed, the smallest percentage that we found for the Thomas-Reiche-Kuhn sum rule was  
%from $168.3\%$ for $\nuc{Ca}{40}$ based on the SRG interaction to $223.8\%$ for $\nuc{Pb}{208}$ based on the UCOM(SRG) interaction.
$168.3\%$, for $\nuc{Ca}{40}$ based on the SRG interaction. 
% to $223.8\%$ for $\nuc{Pb}{208}$ based on the UCOM(SRG) interaction.
%%Deviations of the classical sum rules may also reflect to some extent the degree of convergence of our calculations with respect to the employed single-particle basis and minor inaccuracies due to numerical integrations over a radial grid when using a density-dependent interaction.
Finally, we have found that the energy of the spurious dipole state never exceeds 20keV in the present cases. 
%Our implementation has been checked also by looking at the spurious dipole state due to the restoration of translational invariance. In most cases its energy amounts to at most a few $\keV$, only in one case it reaches $20\keV$.

%%%%%%%%%%%%%%%%%%%%%%%%%%%%%%%%%%%%%%%%%%%%%%%%%%%%%%%%%%%
\begin{figure*}[t]
\centering
\includegraphics*[width=\textwidth]{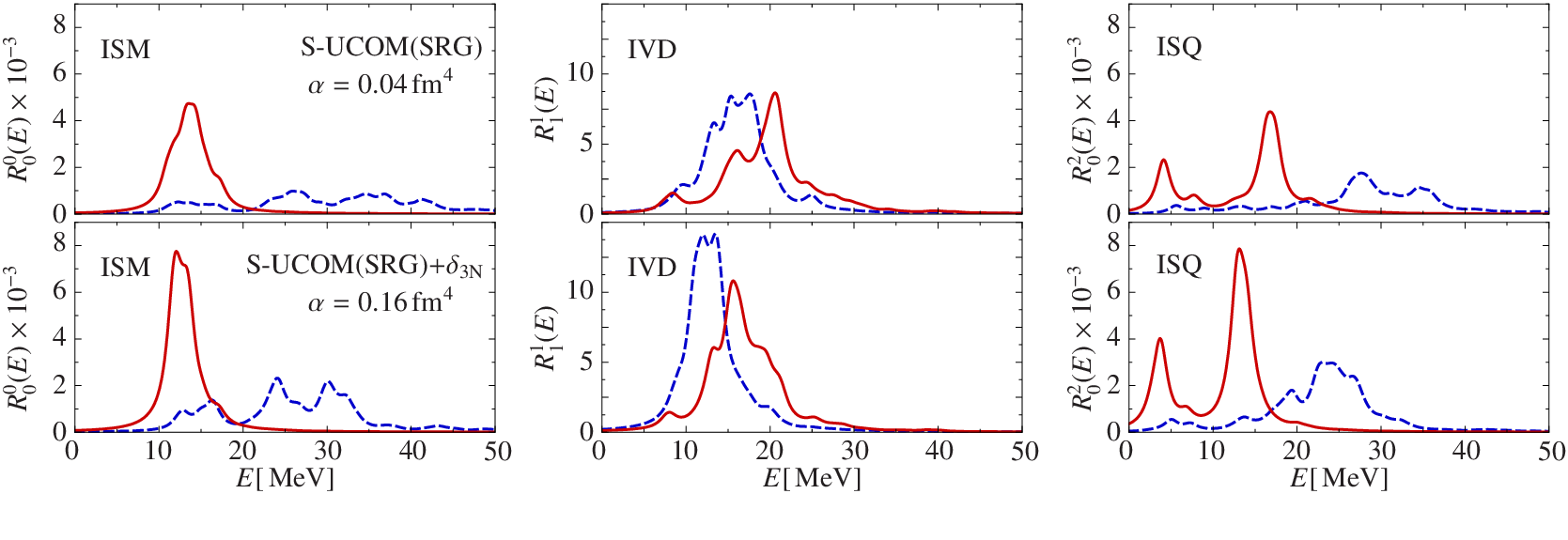}
\vspace{-1cm}
\caption{(Color online) Comparison of HF (\linemediumdashed[FGBlue])\ and RPA (\linemediumsolid[FGRed])\ response functions calculated with the pure two-body S-UCOM(SRG) interaction, (a)-(c), and the two-body S-UCOM(SRG) plus three-body contact interaction, (d)-(f), for $\nuc{Pb}{208}$. The ISM, IVD, and ISQ response functions are given in units of $10^3\fm^4/\MeV$, $e^2\fm^2/\MeV$, and $10^3e^2\fm^4/\MeV$, respectively.}
\label{fig:hf_rpa}
\end{figure*}
%%%%%%%%%%%%%%%%%%%%%%%%%%%%%%%%%%%%%%%%%%%%%%%%%%%%%%%%%%%

In Figure~\ref{fig:hf_rpa} we compare the HF and RPA response functions for all three excitation modes calculated with the S-UCOM(SRG) interaction for $\nuc{Pb}{208}$. The response functions are obtained via a convolution of the calculated discrete strength distribution with a Lorentzian function with a width of $2\MeV$. Fig.~\ref{fig:hf_rpa}(a)-(c) 
%The upper row 
shows the response functions obtained with the pure two-body S-UCOM(SRG) interaction using the flow parameter $\alpha=0.04\fm^4$ while the response functions depicted in 
Fig.~\ref{fig:hf_rpa}(d)-(f) 
%the lower row 
were obtained with the two-body S-UCOM(SRG) interaction plus the three-body contact interaction with $\alpha=0.16\fm^4$ and $\Cnnn=2.2\GeV\fm^6$ (cf. Tab.~\ref{tab:optparametersets}).
The HF response is spread wide in case of both isoscalar modes while it is rather compressed for the isovector excitation. In comparison, the RPA response is compressed significantly and shifted to lower excitation energies in the isoscalar channels which leads to strongly collective excitation modes, the giant resonances. In case of the ISQ mode one observes the excitation of a low-lying $2^+$ state in addition to the giant resonance. The RPA response of the IVD excitation is shifted to higher excitation energies with respect to the HF response, i.e. the residual interaction is attractive in the IS channels and repulsive in the IV channel.

Comparing the HF response functions obtained with the pure two-body interaction with those resulting from the two- plus three-body interaction reveals that the inclusion of the three-body interaction leads to a compression of the response for all considered excitation modes. This compression can be understood by considering the HF single-particle spectra which are spread wide when calculated with a pure two-body interaction (see also \cite{Bia2014}). The repulsion of the three-body interaction increases the level density and thus leads to a compression of the HF response.

The RPA response calculated with the two- plus three-body interaction is concentrated in a narrower resonance structure compared to the response functions obtained with the pure two-body interaction for all three excitation modes. Furthermore, the centroids are shifted to lower energies: the ISM centroid is shifted by $1\MeV$ while the IVD and ISQ centroids are moved by $3\MeV$, respectively. An important mechanism lowering the energies is the compression of the HF spectra, as discussed also in Ref.~\cite{Hergert:2011eh}. The overall effect seems weaker for the monopole resonance. A possible partial explanation is that the addition of the three-body term accompanies a modification of the flow parameter and therefore the non-local terms of the two-body interaction. The latter may affect the IVD and ISQ resonances more strongly than the three body term, while a more balanced effect is at work in the case of the compression mode.

%%%%%%%%%%%%%%%%%%%%%%%%%%%%%%%%%%%%%%%%%%%%%%%%%%%%%%%%%%%
\begin{figure*}[t]
\centering
\includegraphics*[width=\textwidth]{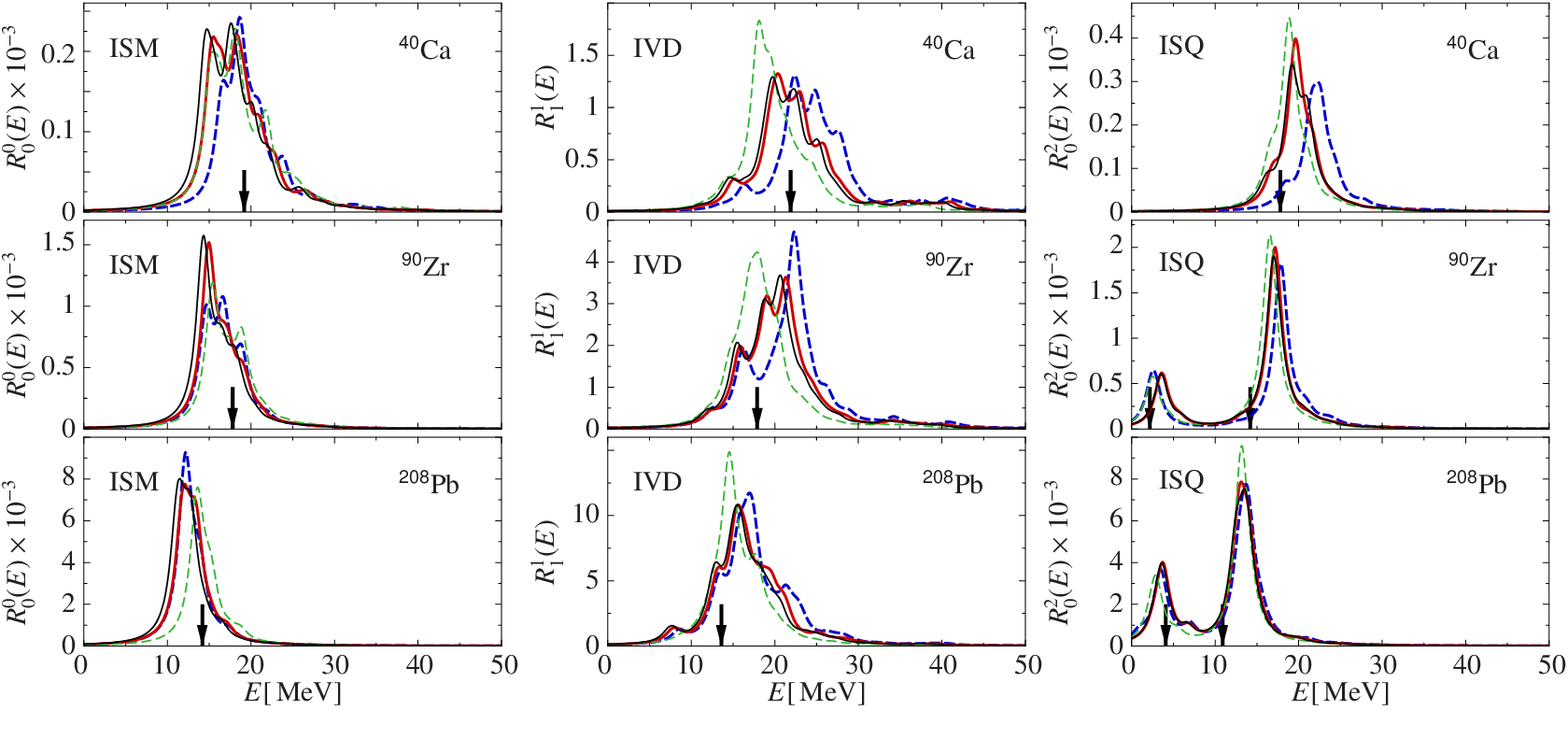}
\vspace{-1cm}
\caption{(Color online) Comparison of giant resonances calculated with the different two- plus three-body interactions:
UCOM(SRG) (\linemediumdashed[FGBlue]), S-UCOM(SRG) (\linemediumsolid[FGRed]), SRG (\linethindashed[FGGreen]), and S-SRG (\linethinsolid[FGBlack]). The ISM, IVD, and ISQ response functions are given in units of $10^3\fm^4/\MeV$, $e^2\fm^2/\MeV$, and $10^3e^2\fm^4/\MeV$, respectively. 
Arrows indicate experimentally extracted centroid energies~\cite{Youngblood:2001mq,Youngblood:1999zza,Veyssiere1974513,Berman:1967zz,Veyssiere1970561,Bertrand1979198,Raman:1201zz,Heisenberg:1982zz}. 
%Centroid energies extracted from experimental data  
%\cite{Youngblood:2001mq,Youngblood:1999zza,Veyssiere1974513,Berman:1967zz,Veyssiere1970561,Bertrand1979198,Raman:1201zz,Heisenberg:1982zz} 
%\cite{Youngblood:2001mq,Youngblood:1999zza,Vey1974,Berman:1967zz,Vey1970,Ber1979,Raman:1201zz,Heisenberg:1982zz} 
%are indicated by arrows.
}
\label{fig:rpa}
\end{figure*}
%%%%%%%%%%%%%%%%%%%%%%%%%%%%%%%%%%%%%%%%%%%%%%%%%%%%%%%%%%%

%%%%%%%%%%%%%%%%%%%%%%%%%%%%%%%%%%%%%%%%%%%%%%%%%%%%%%%%%%%
\begin{table}
\begin{center}
\begin{tabular}{|cc|c|c|c|c|c|c|}
\hline
&& (a) & (b) & (c) & (d) & Exp. & (e) \\
\hline
ISM & $\nuc{Ca}{40}$ & 19.81 & 18.57 & 18.91 & 17.92 & 19.18 & 21.06\\
& $\nuc{Zr}{90}$ & 16.73 & 16.40 & 17.37 & 15.81 & 17.81 & 17.53\\
& $\nuc{Pb}{208}$ & 12.88 & 12.93 & 14.41 & 12.35 & 14.18 & 13.19\\
\hline
IVD & $\nuc{Ca}{40}$ & 25.86 & 23.22 & 20.34 & 22.61 & 21.9 & 22.70\\
& $\nuc{Zr}{90}$ & 22.20 & 20.79 & 18.46 & 20.27 & 17.9 & 19.82\\
& $\nuc{Pb}{208}$ & 17.88 & 16.96 & 15.72 & 16.44 & 13.6 & 16.21\\
\hline
ISQ & $\nuc{Ca}{40}$ & 22.54 & 20.06 & 19.02 & 19.92 & 17.8 & 17.69\\
& $\nuc{Zr}{90}$ & 18.49 & 17.45 & 16.75 & 17.38 & 14.2 & 12.70\\
& $\nuc{Pb}{208}$ & 14.07 & 13.66 & 13.43 & 13.54 & 10.9 & 9.36\\
\hline
\end{tabular}
\end{center}
\vspace{-0.5cm}
\caption{Centroid energies in $\MeV$ obtained with the (a) UCOM(SRG), (b) S-UCOM(SRG), (c) SRG, and (d) S-SRG interactions, compared to experimental values
\cite{Youngblood:2001mq,Youngblood:1999zza,Veyssiere1974513,Berman:1967zz,Veyssiere1970561,Bertrand1979198} and the (e) Gogny D1S interaction.}
\label{tab:centroids}
\end{table}
%%%%%%%%%%%%%%%%%%%%%%%%%%%%%%%%%%%%%%%%%%%%%%%%%%%%%%%%%%%

Figure~\ref{fig:rpa} summarizes the response functions that were obtained with the four different two- plus three-body interactions for $\nuc{Ca}{40}$, $\nuc{Zr}{90}$, and $\nuc{Pb}{208}$ including centroid energies extracted from experiment. All four interactions yield comparable results with only minor differences and are in reasonable agreement with the experimental centroids. 
The energies of the giant dipole and quadrupole resonances tend to be overestimated, but much less so than with the pure two-body UCOM interaction \cite{Paar:2006ua}. 
The systematic nature of these deviations hint at the role of higher-order configurations beyond RPA~\cite{DNS1990,Papakonstantinou:2009zz}.  We shall return to this issue in Sec.~\ref{sec:conclusion}.

The calculated centroid energies are compared to the experimental values and to calculations based on the Gogny D1S interaction in Table~\ref{tab:centroids}. 
In most of the 36 cases the centroids are within $20\%$ of the experimental value. 
The SRG interaction performs particularly well, while the largest deviations are observed for the UCOM(SRG). 
Our results are similar in quality with existing results using phenomenological energy density functionals, e.g. the Gogny D1S interaction listed in Table~\ref{tab:centroids}. 

The results obtained with the S-UCOM(SRG) and the S-SRG interactions are very similar in all aspects confirming the similarities between these two interactions that were already observed earlier \cite{Gunther:2010cs}. Remarkably, the inclusion of a simplistic contact three-body force suffices to cure the pathologies observed in the results with the two-body SRG interaction~\cite{Gunther:2010cs} and to produce  good results. The strength of the three-nucleon term for this interaction is larger than, but of the same order of magnitude as, the strength of the corresponding term for the other three interactions and, even so, weak compared with three-nucleon or density-dependent terms accompanying phenomenological functionals, as discussed in Ref.~\cite{Hergert:2011eh}.

%%%%%%%%%%%%%%%%%%%%%%%%%%%%%%%%%%%%%%%%%%%%%%%%%%%%%%%%%%%%%%%%%%%%%%
%%%%%%%%%%%%%%%%%%%%%%%%%%%%%%%%%%%%%%%%%%%%%%%%%%%%%%%%%%%%%%%%%%%%%%
\section{Discussion and Outlook 
\label{sec:conclusion}} 

The good overall agreement of the calculated response with experiment is obtained despite the systematic deviation of the HF energy (without MBPT corrections) from the experimental binding energy. 
The same remarkable observation was made in \cite{Hergert:2011eh}. 
It is well known that an excellent description of nuclear energies by phenomenological energy density functionals based on HF does not guarantee a good description of collective states within RPA. 
For example, they usually require a high nucleon effective mass, which leads to a strong underestimation of the IVD energy. 
Our present and previous results \cite{Hergert:2011eh} consistently confirm that the HF or HF-Bogolyubov ground-state energy has little to do with the dynamical behavior of the system under study, as described by RPA or quasi-particle RPA.
%We note that the energies of the real RPA ground state and the HF reference state can be significantly different~\cite{Barbieri:2006vf}. 
Instead it is more important to obtain a good description of the radii and the ground-state energies including long-range correlations, e.g. in many-body perturbation theory or RPA. This is a point worth elaborating on. 

It can be argued, that the RPA approach is inconsistent if the HF reference state does not describe well the ground-state properties. 
It has been observed, indeed,  that the energies of the RPA ground state and the HF reference state can be significantly different~\cite{Barbieri:2006vf}. 
In this respect we note that 1) explicit use of a correlated RPA reference state has a rather weak effect on the description of giant resonances~\cite{PRP2007} and that 2) it is in a sense a more consistent approach, if the interaction used describes well the ground states within a (converged) perturbative and beyond-mean-field approach. We point out that phenomenological interactions which describe well nuclei within HF would fail badly in extended many-body theories. 

Ideally, a correlated, beyond-mean-field wavefunction should be used as a reference state and excitation configurations should be built consistently on such a correlated ground state, taking the depletion of the Fermi sea into account~\cite{Row1968a}.  
Nontheless, it is in line with the philosophy of the equations-of-motion method to employ an approximate ground state (e.g., HF), conditionally. The condition is that the expectation values of the relevant low-rank operators (in the many-body sense), resulting from (double) commutators of higher-rank ones, not be sensitive to correlations. The relevant operators in the present cases pertain to long-range spatial properties like radii, which, indeed, are insensitive to correlations. 

First-order RPA need not be the final converged result. 
It is expected that higher-order configurations will shift the resonances to lower energies. We have seen that the energies of the major giant resonances are overestimated in the present approach. 
It is therefore a very interesting possibility that, for example, second-order RPA (SRPA) can provide more converged and accurate results. 
%This possibility shall be examined in the future. 
There are two issues to be addressed in this respect: First, SRPA based on an uncorrelated reference state is inconsistent and leads to instabilities~\cite{Pap201X}; nonetheless, it has been demonstrated that the SRPA results on IVD and ISQ giant resonances are not sensitive to the treatment of ground-state correlations~\cite{PaR2010,Bia2014} (contrary to low-lying states). Second, it remains to be seen whether the three-body term, which would introduce a large number of non-zero matrix elements in the B matrix of SRPA~\cite{GGC2011a}, can moderate the resonance shift, as compared to the large shifts observed using a two-body interaction. 
Clearly the above issues and possibilities are worth exploring further. 

In conclusion, we have investigated giant resonances using unitarily transformed two-nucleon interactions supplemented by a simple phenomenological three-body contact interaction. For the three considered nuclei, $\nuc{Ca}{40}$, $\nuc{Zr}{90}$, and $\nuc{Pb}{208}$ we achieved a reasonable agreement with experimental centroid energies, where the SRG interaction yielded the smallest deviations. 
A reasonable description of charge radii was found important to obtain a good starting point for RPA calculations.
The present study represents 
an intermediate step towards a consistent inclusion of chiral two- plus three-nucleon interactions for the study of collective excitations  
with various methods including ground state correlations or higher-order configurations. 
Following this benchmark, 
%the present hybrid interactions can be applied for the quantitative description of other collective phenomena in closed and open-shell nuclei. 
several studies are in progress beyond HF-RPA, aiming at the quantitative description of collective phenomena in closed and open-shell nuclei. 
Apart from computational challenges, the systematic reproduction of ground-state radii will be a key issue when working with such Hamiltonians.

%%%%%%%%%%%%%%%%%%%%%%%%%%%%%%%%%%%%%%%%%%%%%%%%%%%%%%%%%%%%%%%%%%%%%%
%%%%%%%%%%%%%%%%%%%%%%%%%%%%%%%%%%%%%%%%%%%%%%%%%%%%%%%%%%%%%%%%%%%%%%
%%%%%%%%%%%%%%%%%%%%%%%%%%%%%%%%%%%%%%%%%%%%%%%%%%%%%%%%%%%%%%%%%%%%%%
\section*{Acknowledgments}

This work is supported by the Deutsche Forschungsgemeinschaft through contract SFB 634, by the Helmholtz International Centre for FAIR (HIC for FAIR) within the framework of the LOEWE program launched by the State of Hesse, by the BMBF through contracts 06DA9040I, 06DA7047I, by the BMBF-FSP 302 ``NUSTAR.de'',  
by the ANR project ``SN2NS'' 
and by the Rare Isotope Science Project of the Institute for Basic Science funded by the Ministry of Science, ICT and Future Planning and the National Research Foundation of Korea (2013M7A1A1075766). 

%%%%%%%%%%%%%%%%%%%%%%%%%%%%%%%%%%%%%%%%%%%%%%%%%%%%%%%%%%%%%%%%%%%%%%
%%%%%%%%%%%%%%%%%%%%%%%%%%%%%%%%%%%%%%%%%%%%%%%%%%%%%%%%%%%%%%%%%%%%%%
%%%%%%%%%%%%%%%%%%%%%%%%%%%%%%%%%%%%%%%%%%%%%%%%%%%%%%%%%%%%%%%%%%%%%%

%\bibliography{bibrpa} 

%\end{document}  

\end{document}